\documentclass[a4paper, 11pt]{article}

\usepackage[cp1251]{inputenc}
\usepackage[russian]{babel}

\usepackage{amsfonts,amsmath,amssymb,amsthm,latexsym}
\usepackage{eufrak,ifthen}

\newcommand{\ArticleName}[1]{\renewcommand{\theArticleName}{#1}\vspace{-2mm}\par\noindent {\LARGE\bf  #1\par}}
\newcommand{\Author}[1]{\vspace{5mm}\par\noindent {\it #1} \par\vspace{2mm}\par}
\newcommand{\Address}[1]{\vspace{2mm}\par\noindent {\it #1} \par}
\newcommand{\Email}[1]{\ifthenelse{\equal{#1}{}}{}{\par\noindent {\rm E-mail: }{\it  #1} \par}}
\newcommand{\URLaddress}[1]{\ifthenelse{\equal{#1}{}}{}{\par\noindent {\rm URL: }{\tt  #1} \par}}
\newcommand{\EmailD}[1]{\ifthenelse{\equal{#1}{}}{}{\par\noindent {$\phantom{\dag}$~\rm E-mail: }{\it  #1} \par}}
\newcommand{\URLaddressD}[1]{\ifthenelse{\equal{#1}{}}{}{\par\noindent {$\phantom{\dag}$~\rm URL: }{\tt  #1} \par}}

\newcommand{\Abstract}[1]{\vspace{6mm}\par\noindent\hspace*{10mm}
\parbox{140mm}{\small {\bf Abstract.} #1}\par}

\newcommand{\Keywords}[1]{\vspace{3mm}\par\noindent\hspace*{10mm}
\parbox{140mm}{\small {\it Key words:} \rm #1}\par}
\newcommand{\Classification}[1]{\vspace{3mm}\par\noindent\hspace*{10mm}
\parbox{140mm}{\small {\it 2000 Mathematics Subject Classification:} \rm #1}\vspace{3mm}\par}

\newcommand{\theArticleName}{Article name}

\newcommand{\abs}[1]{\left\vert#1\right\vert}
\newtheorem{theorem}{Theorem}

{\theoremstyle{definition}

\begin {document}

\ArticleName{On the solutions of the nonlinear Liouville hierarchy
  \footnote{This work was partially supported by the WTZ grant}}

\Author{V.~O. SHTYK}

\Address{Institute of Mathematics of NAS of Ukraine, 3
         Tereshchenkivs'ka Str., 01601 Kyiv-4, Ukraine}
\EmailD{shtyk@imath.kiev.ua}

{\Abstract%[Анотація]}

      We investigate the initial-value problem of the non-linear
      Liouville hierarchy. For the general form of the interaction
      potential we construct an explicit solution in terms of
      an expansion over particle clusters whose evolution is
      described by the corresponding-order cumulant of
      evolution operators of a system of finitely many particles.
      For the
      initial data from the space of integrable functions the
      existence of a strong solution of the Cauchy
      problem is proved.


\Classification{82C05; 37A60.}

\section{Introduction}
The nonlinear Liouville hierarchy that describes the evolution of
correlation functions, arises in many problems of statistical
mechanics concerning many-particle systems
\cite{1BJinr,2Green,7B_Lec_49}. However, today it is still
insufficiently studied from the
mathematical point of view.\\
\indent It is well known \cite{4CGP97}, that all possible states of
a classical system of a finite number of particles are described by
the functions interpreted  as probability density functions. These
functions are solutions of the initial-value problem of the
Liouville hierarchy
--- the first-order partial
differential equations, whose characteristic equations are Hamilton
equations. If the state of a system is presented in terms of a
cluster expansion in new (correlation) functions one evidently
obtains an equivalent description of this state. Now evolution of
the correlation functions is determined by the nonlinear Liouville
hierarchy
--- certain  nonlinear first-order partial differential
equations.\\
\indent In this paper an explicit solution of such nonlinear
equations is constructed and presented as an expansion in terms of
particle clusters whose evolution is described by a cumulant
(semi-invariant) of the evolution operators. The latter are
determined by the solutions of the characteristic equations of the
linear Liouville equation, i.e., the Hamilton equations. The
interaction potential of the general form is considered, which makes
it possible to describe the general structure of the generator of
the nonlinear Liouville hierarchy. The existence of a strong
solution of the Cauchy problem with initial data from the space of
integrable functions is proved. A formal treatment of the nonlinear
Liouville hierarchy for the case of a pairwise interaction
potential,  was given by
Bogolyubov and Green \cite{1BJinr,2Green}.\\
    \indent It should be noted that the nonlinear Liouville hierarchy
    is basic in the substantiation of the derivation of the nonlinear Bogolyubov hierarchy \cite{2Green}
      whose solutions describe the correlation dynamics of
      infinite systems of particles. Moreover, this concerns the mathematical
       substantiation of the correlation-weakening principle.
The correlation functions may be employed to directly calculate the
specific characteristics of the system, i.e., fluctuations, defined
as the average values of the square deviations of an observable from
its average value, as well as macroscopic values which are not
averages of observables. The construction of the nonlinear
Bogolyubov hierarchy and the analysis of the solutions thereof for
on the basis of the results obtained here will be given in an other
paper.
\section{Initial-Value Problem of Nonlinear Liouville Hierarchy}
Let us consider a system of non-fixed number of identical particles
with phase coordinates $x_{i}=(q_{i},p_{i})\in\mathbb{R}^{\nu
}\times\mathbb{R}^{\nu}$, $\nu\geq 1$. The relevant Hamiltonian is
given by the formula
\begin{center}$H_{n}=\sum\limits_{i=1}^{n}\frac{p_{i}^{2}}{2}+\sum\limits_{k=1}^{n}\sum\limits_{i_{1}<i_{2}<\ldots<i_{k}=1}^{n}
    \Phi_{k}(q_{i_{1}},q_{i_{2}},\ldots,q_{i_{k}}),$
\end{center}
where $\Phi_{k}$ is the $k$-th-order interaction potential. In what
follows we assume that the interaction potential $\Phi_{k}$,
$k\geq1$, satisfies the necessary conditions which provide the
existence
of global in time solutions of the Hamilton equations.\\
\indent The state of the system can be described by a sequence
$g(t)=(0,g_{1}(t,x_1),\ldots\\\ldots,g_{n}(t,x_{1},\ldots,x_{n}),\ldots)$,
of correlation functions $g_{n}(t,x_{1},\ldots,x_{n})$ defined on
the phase space $\mathbb{R}^{\nu n}\times\mathbb{R}^{\nu n}$,
$\nu\geq 1$ and symmetric with respect to the  permutations of
arguments
$x_{1},\ldots,x_{n}.$ \\
\indent The evolution of the states of the above system
 is described by
the initial-value problem of the  nonlinear Liouville hierarchy
\begin{eqnarray}\label{1}
    &&\frac{d}{dt}g_{n}(t,Y)=(-\mathcal{L}_{|Y|}(Y))g_{n}(t,Y)\,+\nonumber\\
    &&\qquad\quad+\sum\limits_{\substack{{\texttt{P}:\,Y=\bigcup\limits_iX_i}\\|\texttt{P}|>1}}
    \sum\limits_{\substack{Z_{i}\subset X_{i}}}
    \big(-\mathcal{L}^{int}_{|\bigcup\limits_{i=1}^{|\texttt{P}|}Z_{i}|}
    \big(\bigcup\limits_{i=1}^{|\texttt{P}|}Z_{i}\big)\big)\prod_{X_i\subset \texttt{P}}g_{\abs
    {X_i}}(t,X_i),
    \end{eqnarray}
\begin{equation}\label{2}
\qquad g_{n}(t,Y)\big|_{t=0}=g_{n}(0,Y),\qquad n\geq1,
\end{equation}
where the following notation is used $Y\equiv(x_{1},\ldots,x_{n})$,
$|Y|=n$  denotes the number of elements of the set $Y$,
$\sum\limits_{\texttt{P}:\,Y=\bigcup\limits_iX_i}$ is the sum over
all possible (in this case) decompositions $\texttt{P}$  of the set
$Y$ into $|\texttt{P}|$ nonempty mutually disjoint subsets,
\,$\sum\limits_{\substack{Z_{i}\subset X_{i}}}$
 - is the sum over all nonempty subsets $Z_{i}\subset X_{i}$.
The Liouville operator $\mathcal{L}_{n}$ for the Hamiltonian $H_{n}$
 is described by the formulas
      \begin{eqnarray}\label{liyvilnian}
      &&\mathcal{L}_{n}\equiv\mathcal{L}_{n}(x_{1},\ldots,x_{n})=
\sum\limits_{i=1}^{n}\langle
p_i,\frac{\partial}{\partial{q_j}}\rangle+
\sum\limits_{k=2}^{n}\sum\limits_{i_{1}<i_{2}<\ldots<i_{k}=1}^{n}
\mathcal{L}_{k}^{int}(x_{i_{1}},x_{i_{2}},\ldots,x_{i_{k}}),\\
&&\mathcal{L}_{k}^{int}(x_{i_{1}},x_{i_{2}},\ldots,x_{i_{k}})
=-\sum\limits_{j=1}^{n}\langle\frac{\partial}{\partial{q_j}}
\Phi_{k}(q_{i_{1}},q_{i_{2}},\ldots,q_{i_{k}}),\frac{\partial}{\partial{p_j}}\rangle,\nonumber
       \end{eqnarray}
where $\langle\cdot,\cdot \rangle$ - is the scalar product.\\
 \indent The simplest examples of the nonlinear Liouville hierarchy are given by:
    \begin{eqnarray*}
    &&\frac{d}{dt}g_{1}(t,x_1)=-\mathcal{L}_{1}(x_1)g_{1}(t,x_1),\nonumber\\
       &&\frac{d}{dt}g_{2}(t,x_1,x_2)=-\mathcal{L}_{2}(x_1,x_2)g_{2}(t,x_1,x_2)-\mathcal{L}^{int}_{2}(x_1,x_2)g_{1}(t,x_1)g_{1}(t,x_2),\nonumber\\
       &&\frac{d}{dt}g_{3}(t,x_1,x_2,x_3)=-\mathcal{L}_{3}(x_1,x_2,x_3)g_{3}(t,x_1,x_2,x_3)+\nonumber\\
    &&\quad+\big(-\mathcal{L}^{int}_{2}(x_1,x_2)-\mathcal{L}^{int}_{2}(x_1,x_3)-\mathcal{L}^{int}_{3}(x_1,x_2,x_3)\big)
    g_{1}(t,x_1)g_{2}(t,x_2,x_3)+ \nonumber\\
    &&\quad+\big(-\mathcal{L}^{int}_{2}(x_1,x_2)-\mathcal{L}^{int}_{2}(x_2,x_3)-\mathcal{L}^{int}_{3}(x_1,x_2,x_3)\big)
    g_{1}(t,x_2)g_{2}(t,x_1,x_3)+ \nonumber\\
    &&\quad+\big(-\mathcal{L}^{int}_{2}(x_1,x_3)-\mathcal{L}^{int}_{2}(x_2,x_3)-\mathcal{L}^{int}_{3}(x_1,x_2,x_3)\big)
    g_{1}(t,x_3)g_{2}(t,x_1,x_2)+ \nonumber\\
    &&\quad-\mathcal{L}^{int}_{3}(x_1,x_2,x_3)g_{1}(t,x_1)g_{1}(t,x_2)g_{1}(t,x_3). \nonumber
\end{eqnarray*}
 \indent We note that in the case of a pairwise interaction potential,
 $(k=2)$,
 the nonlinear Liouville hierarchy (\ref{1}) is simpler. For example,
  expression for $g_{3}(t)$ does not contain members with
   $\mathcal{L}^{int}_{3}$. This case was
  considered by Green \cite{2Green}.\\
 \indent In this work the solution (\ref{1})-(\ref{2})
is shown to be given by the formula
    \begin{equation}\label{rozv}
     g_{n}(t,Y)=\sum\limits_{\texttt{P}:\,Y=\bigcup\limits_iX_i}
        \mathfrak{A}_{|\texttt{P}|}(t,Y_{X_{i}})
        \prod_{X_i\subset \texttt{P}}g_{\abs {X_i}}(0,X_i),
           \end{equation}
    where $Y=(x_{1},\ldots,x_{n})$,
    $Y_{X_{i}}\equiv(X_{1},\ldots,X_{|\texttt{P}|})$,
     $\sum\limits_{\texttt{P}:\,Y=\bigcup\limits_iX_i}$ ---
    the sum of all possible (in this case) decomposition $\texttt{P}$
of the set $Y$ into $|\texttt{P}|$ nonempty mutually disjoint
subsets $X_i$.
    The evolution operator $\mathfrak{A}_{|\texttt{P}|}(t)$
     i.e., the cumulant (semi-invariant) of the order $|\texttt{P}|$ is
     given
     by the expression \cite{3GR02,5GRS04}
     \begin{equation}\label{kymyl}
     \mathfrak{A}_{|\texttt{P}|}(t,Y_{X_{i}})=
     \sum\limits_{\texttt{P}^{'}:\,Y_{X_{i}}=\bigcup\limits_kZ_k}
     (-1)^{|\texttt{P}^{'}|-1}(|\texttt{P}^{'}|-1)!
        \prod_{Z_k\subset \texttt{P}^{'}}S_{\abs {Z_k}}(-t,Z_k).
    \end{equation}
 The evolution operators $S_{n}(-t),\, n\geq 1$, are given by
\begin{eqnarray}\label{S(t)}
\big(S_{n}(-t)g_{n}(0)\big)(x_{1},\ldots,x_{n})=
    g_{n}\big(0,X_{1}(-t,x_{1},\ldots,x_{n}),\ldots,X_{n}(-t,x_{1},\ldots,x_{n})\big),
\end{eqnarray}
where $\{X_{i}(-t,x_{1},\ldots,x_{n})\}_{i=1}^{n}$ is the solution
of the relevant initial-value problem for the Hamilton equations.
The properties of a group of evolution operator (\ref{S(t)}) are
described in \cite{6PGM02}.
   \\ \indent Let us consider the simplest examples
of expansions (\ref{rozv}) with the following notation: the argument
$x_{i}\cup x_{j}$ implies, that two particles $i$-th and $j$-th
evolve as a cluster. Thus, if the arguments of the operator are
clusters, they enter on equal terms the expansions in series of the
evolution operators $S_{n}(-t)$, the order of the cumulant being
equal to the number of its cluster arguments .
  \begin{eqnarray*}
    &&g_{1}(t,x_1)=\mathfrak{A}_{1}(t,x_1)g_{1}(0,x_1),\\
    &&g_{2}(t,x_1,x_2)=\mathfrak{A}_{1}(t,x_1\cup x_2)g_{2}(0,x_1,x_2)+\mathfrak{A}_{2}(t,x_1,x_2)g_{1}(0,x_1)g_{1}(0,x_2),\\
    &&g_{3}(t,x_1,x_2,x_3)= \mathfrak{A}_{1}(t,x_1\cup x_2\cup x_3)g_{3}(0,x_1,x_2,x_3)+
    \mathfrak{A}_{2}(t,x_1,x_2\cup
    x_3)g_{1}(0,x_1)g_{2}(0,x_2,x_3)+\\
    &&\quad+\mathfrak{A}_{2}(t,x_1\cup x_3,x_2)g_{1}(0,x_2)g_{2}(0,x_1,x_3)
    +\mathfrak{A}_{2}(t,x_1\cup
    x_2,x_3)g_{1}(0,x_3)g_{2}(0,x_1,x_2)+\\
    &&\quad+\mathfrak{A}_{3}(t,x_1,x_2,x_3)g_{1}(0,x_1)g_{1}(0,x_2)g_{1}(0,x_3),
\end{eqnarray*}
where, for example, the cumulants (\ref{kymyl}) are given by
\begin{eqnarray*}
    &&\mathfrak{A}_{2}(t,x_1\cup
    x_2,x_3)=S_{3}(-t,x_1,x_2,x_3)-S_{1}(-t,x_3)S_{2}(-t,x_1,x_2)\\
    &&\mathfrak{A}_{3}(t,x_1,x_2,x_3)=S_{3}(-t,x_1,x_2,x_3)-\\
    &&\quad-S_{1}(-t,x_1)S_{2}(-t,x_2,x_3)-S_{1}(-t,x_2)S_{2}(-t,x_1,x_3)-S_{1}(-t,x_3)S_{2}(-t,x_1,x_2)+\\
    &&\quad+2!S_{1}(-t,x_1)S_{1}(-t,x_2)S_{1}(-t,x_3).
\end{eqnarray*}
 \indent Formally, the nonlinear Liouville hierarchy (\ref{1}) can be derived from the
  sequence of (linear) Liouville equations which describe the evolution of all possible
   states of the system of non-fixed number of particles (the sequence
  $D(t)=(1, D_{1}(x_{1}),\ldots,D_{n}(x_{1},\ldots,x_{n}),\ldots)$,
  with the function $D_{n}(t)$ being regarded as the density of probability distribution
  of the
  $n$-particles system) \cite{4CGP97}
         \begin{eqnarray*}
         && \frac{\partial D_{n}(t)}{\partial
         t}=-\mathcal{L}_{n}D_{n}(t),\\
         &&D_{n}(t)\big|_{t=0}=D_{n}(0),\quad n\geq1,
         \end{eqnarray*}
            provided the state of the system is described in terms of correlation
            functions, i.e.,
         \begin{equation}\label{g_(D)}
        g_{n}(t,Y)=
        \sum\limits_{\texttt{P}:\,Y=\bigcup\limits_iX_i}(-1)^{\abs {\texttt{P}}-1}(\abs {\texttt{P}} -1)!\,
        \prod_{X_i\subset \texttt{P}}D_{\abs {X_i}}(t,X_i),
        \end{equation}
or at the initial time instant
        \begin{equation}\label{g(D)}
        g_{n}(0,Y)=
        \sum\limits_{\texttt{P}:\,Y=\bigcup\limits_iX_i}
        (-1)^{\abs {\texttt{P}}-1}(\abs {\texttt{P}} -1)!\,
          \prod_{X_i\subset \texttt{P}}D_{\abs {X_i}}(0,X_i),\qquad n\geq 1,
         \end{equation}
           where  $Y\equiv(x_{1},\ldots,x_{n})$,
             $\sum\limits_\texttt{P}$ is the sum over all decompositions  $\texttt{P}$
of the set $Y$ into $|\texttt{P}|$ nonempty mutually disjoint
subsets $X_{i}$.\\
For example,
     \begin{eqnarray}
     &&g_{1}(t,x_1)=D_{1}(t,x_1),\nonumber\\
     &&g_{2}(t,x_1,x_2)=D_{2}(t,x_1,x_2)-D_{1}(t,x_1)D_{1}(t,x_2).\nonumber
     \end{eqnarray}
\indent The solution (\ref{rozv}) of the initial-value problem
(\ref{1})-(\ref{2}) can be formally derived from(\ref{g_(D)}) and
(\ref{g(D)}) provided one takes into account that, within the
context of (\ref{g(D)}), we have
        \begin{eqnarray*}
        D_{|X_{i}|}(0,X_{i})=
        \sum\limits_{\texttt{P}_{2}:\,X_{i}=\bigcup\limits_kZ_k}
        \prod_{Z_k\subset \texttt{P}_{2}}g_{\abs {Z_k}}(0,Z_k).
           \end{eqnarray*}
Then, inasmuch as the solution of the Liouville equation is given by
         \begin{equation*}
         D_{n}(t)=S_{n}(-t)D_{n}(0),
         \end{equation*}
         where $S_{n}(-t)$ is determined by the solutions of Hamiltonian
         equations according to  (\ref{S(t)}),
         we have
        \begin{eqnarray}\label{g(g(0))}
     &&g_{n}(t,Y)=\sum\limits_{\texttt{P}_{1}:\,Y=\bigcup\limits_iX_i}
     (-1)^{\abs {\texttt{P}_1}-1}(\abs {\texttt{P}_1} -1)!
        \prod_{X_i\subset \texttt{P}_{1}}\times\nonumber\\
        &&\qquad\qquad\qquad\qquad\qquad\qquad\times S_{\abs {X_i}}(-t,X_i)
        \sum\limits_{\texttt{P}_{2}:\,X_{i}=\bigcup\limits_kZ_k}
        \prod_{Z_k\subset \texttt{P}_{2}}g_{\abs {Z_k}}(0,Z_k).
           \end{eqnarray}
           \indent Having collected in (\ref{g(g(0))}) the terms with similar
           product of functions $g_{\abs
{Z_k}}(0,Z_k)$, one obtains (\ref{rozv}).\\
\section{The existence theorem for the initial-value problem of the nonlinear Liouville hierarchy.}
\indent Suppose $L^{1}_{n}$ is the Banach space of integrable
functions $g_{n}(x_{1},\ldots,x_{n})$, defined on the phase space
$\mathbb{R}^{\nu n}\times\mathbb{R}^{\nu n}$, $\nu\geq 1$ of
$n$-particle system, symmetric under the perturbations of arguments.
The norm of an element   $g_{n}$ of $L^{1}_{n}$ is denoted by
\begin{equation*}
    \|g_{n}\|=\int_{\mathbb{R}^{\nu}\times\mathbb{R}^{\nu}} dx_{1}\ldots dx_{n}
    |g_{n}(x_{1},\ldots,x_{n})|,
   \end{equation*}
   $L^{1}_{n,0}\subset L^{1}_{n}$ is a subspace of continuously
differentiable functions with compact supports.\\
\indent The following theorem is true.

\begin{theorem}  If $g_{n}(0)\in L^{1}_{n,0} \subset L^{1}_{n} $, $n\geq
1$, then for $t\in\mathbb{R}^{1}$ there exists a unique strong
solution to the initial-value problem (\ref{1})-(\ref{2}) of the
nonlinear Liouville hierarchy (\ref{1}) given by
    \begin{equation*}
     g_{n}(t,Y)=\sum\limits_{\texttt{P}:\,Y=\bigcup\limits_iX_i}
        \mathfrak{A}_{|\texttt{P}|}(t,Y_{X_{i}})
        \prod_{X_i\subset \texttt{P}}g_{\abs {X_i}}(0,X_i) ,\nonumber
    \end{equation*} where $\mathfrak{A}_{|\texttt{P}|}(t)$ is a cumulant (semi-invariant)
    of the order $|\texttt{P}|$
    (\ref{kymyl})
     of the evolution operators (\ref{S(t)}).
\end{theorem}

\begin{proof} Let us show that the expansion
(\ref{rozv}) is defined in $L^{1}_{n}$.\\
\indent Indeed, within the context of the corollary of the Liouville
theorem \cite{6PGM02} (isometric property of operators $S_{\abs
{Z_k}}(-t,Z_k)$) the following estimate is valid
     \begin{eqnarray*}
        &&\|g_{n}(t)\|=\int dY|\sum\limits_{\texttt{P}:\,Y=\bigcup\limits_iX_i}
        \mathfrak{A}_{|\texttt{P}|}(t,Y_{X_{i}})
        \prod_{X_i\subset \texttt{P}}g_{\abs {X_i}}(0,X_i)|=\\
        &&=\int dY|\sum\limits_{\texttt{P}:\,Y=\bigcup\limits_iX_i}
             \sum\limits_{\texttt{P}^{'}:\,Y_{X_{i}}=\bigcup\limits_kZ_k}
     (-1)^{|\texttt{P}^{'}|-1}(|\texttt{P}^{'}|-1)!
        \prod_{Z_k\subset \texttt{P}^{'}}S_{\abs {Z_k}}(-t,Z_k)
        \prod_{X_i\subset \texttt{P}}g_{\abs {X_i}}(0,X_i)|\leq\\
        &&\leq \sum\limits_{\texttt{P}:\,Y=\bigcup\limits_iX_i}
             \sum\limits_{\texttt{P}^{'}:\,Y_{X_{i}}=\bigcup\limits_kZ_k}
     (|\texttt{P}^{'}|-1)!
        \prod_{X_i\subset \texttt{P}}\|g_{\abs {X_i}}(0)\|\leq n!e^{n+1}
        \sum\limits_
        {\texttt{P}:\,Y=\bigcup\limits_iX_i}
             \prod_{X_i\subset \texttt{P}}\|g_{\abs {X_i}}(0)\|<\infty,
     \end{eqnarray*}
i.e., $g_{n}(t)\in L^{1}_{n}$ for any $t\in\mathbb{R}^{1}$.\\
\indent Let us prove that the expansion (\ref{rozv}) is a strong
solution of the Cauchy problem of the nonlinear Liouville hierarchy
(\ref{1})-(\ref{2}). \\
To do this we first differentiate the functions $g_{n}(t)$
 with respect to time with regard for the point-by-point
convergence. Let $g_{n}(0)\in L^{1}_{n,0}$, then inasmuch
       \begin{equation*}
       \frac{d}{dt}S_{|X_{i}|}(-t,X_{i})=-\mathcal{L}_{|X_{i}|}(X_{i})S_{|X_{i}|}(-t,X_{i}),
       \end{equation*}
       where the Liouvilian $\mathcal{L}_{|X_{i}|}$ is defined by
(\ref{liyvilnian}), and according to (\ref{g(g(0))}), for each fixed
point $Y$ on any compact from $\mathbb{R}^{\nu
n}\times\mathbb{R}^{\nu n}$ , we have
\begin{eqnarray*}
       &&\frac{d}{dt}g_{n}(t,Y)=\frac{d}{dt}\sum\limits_{\texttt{P}:Y=\bigcup\limits_iX_i}
        \mathfrak{A}_{|\texttt{P}|}(t,Y_{X_{i}})
        \prod_{X_i\subset \texttt{P}_{1}}g_{\abs {X_i}}(0,X_i)=\nonumber\\
        &&\qquad=\frac{d}{dt}\sum\limits_{\texttt{P}_{1}:\,Y=\bigcup\limits_iX_i}
     (-1)^{\abs {\texttt{P}_1}-1}(\abs {\texttt{P}_1}-1)!
        \prod_{X_i\subset \texttt{P}_{1}}S_{\abs {X_i}}(-t,X_i)
        \sum\limits_{\texttt{P}_{2}:\,X_{i}=\bigcup\limits_kZ_k}
        \prod_{Z_k\subset \texttt{P}_{2}}g_{\abs {Z_k}}(0,Z_k)=\nonumber\\
        &&\qquad=\sum\limits_{\texttt{P}_{1}:\,Y=\bigcup\limits_iX_i}
     (-1)^{\abs {\texttt{P}_1}-1}(\abs {\texttt{P}_1} -1)!
        \sum\limits_{X_j\subset P_1}(-\mathcal{L}_{|X_{j}|}(X_j))
        \prod_{X_i\subset \texttt{P}_{1}}S_{\abs {X_i}}(-t,X_i)\times\nonumber\\
        &&\qquad\times\sum\limits_{\texttt{P}_{2}:\,X_{i}=\bigcup\limits_kZ_k}
        \prod_{Z_k\subset \texttt{P}_{2}}g_{\abs {Z_k}}(0,Z_k).\nonumber
\end{eqnarray*}
In view of the equality
        \begin{eqnarray}
        S_{\abs{X_i}}(-t,X_i)\sum\limits_{P_{2}:\,X_{i}=\bigcup\limits_kZ_k}
        \prod_{Z_k\subset \texttt{P}_{2}}g_{\abs {Z_k}}(0,Z_k)=\sum\limits_{\texttt{P}_2:\,X_i=\bigcup\limits_iZ_k}\,
        \prod_{Z_k\subset \texttt{P}_2}g_{\abs {Z_k}}(t,Z_k),\nonumber
        \end{eqnarray}
which follows from $(\ref{g(g(0))})$, we have
        \begin{eqnarray*}
        &&\frac{d}{dt}g_{n}(t,Y)=
        \sum\limits_{\texttt{P}_1:\,Y=\bigcup\limits_iX_i}(-1)^{\abs {\texttt{P}_1}-1}(\abs {\texttt{P}_1} -1)!
        \sum\limits_{X_j\subset \texttt{P}_1}
        (-\mathcal{L}(X_j))\prod_{X_i\subset \texttt{P}_1}\sum\limits_{\texttt{P}_2:\,X_i=\bigcup\limits_iZ_k}\,
        \prod_{Z_k\subset \texttt{P}_2}g_{\abs {Z_k}}(t,Z_k)=\\
        &&\quad=(-\mathcal{L}_{|Y|}(Y))g_{n}(Y)\,+
        \sum\limits_{\substack{{\texttt{P}_1:\,Y=\bigcup\limits_iX_i}\\\texttt{P}_{1}>1}}
        \sum\limits_{\texttt{P}_2:\,Y_{X_{i}}=\bigcup\limits_kZ_k}
        (-1)^{\abs {\texttt{P}_2}-1}(\abs {\texttt{P}_2} -1)!\times\\
        &&\quad\times\sum\limits_{Z_k\subset \texttt{P}_2}
        (-\mathcal{L}_{|Z_{k}|}(Z_k))\prod_{X_i\subset \texttt{P}_1}g_{\abs
        {X_i}}(t,X_i),
        \end{eqnarray*}
the notation being similar to that in (\ref{rozv}) and
 $Y_{X_{i}}\equiv(X_{1},\ldots,X_{|\texttt{P}|})$ is the set whose
  elements are $|\texttt{P}|$  subsets $X_{i}\subset Y$.\\
 \indent Since the following identity holds
 \begin{equation*}
    \sum\limits_{\texttt{P}:\,Y_{X_{i}}=\bigcup\limits_kZ_k}
        (-1)^{\abs {\texttt{P}}-1}(\abs {\texttt{P}}-1)!\sum\limits_{Z_k\subset \texttt{P}}
        (-\mathcal{L}_{|Z_{k}|}(Z_k))=\sum\limits_{\substack{Z_{i}\subset
        X_{i}}}
        \big(-\mathcal{L}^{int}_{|\bigcup\limits_{i=1}^{|\texttt{P}|}Z_{i}|}
        \big(\bigcup\limits_{i=1}^{|\texttt{P}|}Z_{i}\big)\big),
     \end{equation*}
where $\,\sum\limits_{\substack{Z_{i}}\subset X_{i}}$
 is a sum over all nonempty subsets
 $Z_{i}\subset X_{i}$ and operator
$\mathcal{L}^{int}_{|\bigcup\limits_{i=1}^{|\texttt{P}|}Z_{i}|}$ is
defined by formulas (\ref{liyvilnian}), we come to equations
$(\ref{1})$. Thus, formula (\ref{rozv}) determines the solution of
the initial-value problem of the nonlinear Liouville hierarchy
(\ref{1}) from the viewpoint of the
point-by-point convergence.\\
\indent Let us show that the strong derivative of the solution
(\ref{rozv}), reproduces the generator of the nonlinear Liouville
hierarchy
(\ref{1}) in the subspace $L^{1}_{n,0}\subset L^{1}_{n}$ .\\
\indent For $g_{n}(0)\in L^{1}_{n,0}$, in the sense of convergence
norm of the space $L^{1}_{n}$, we have
 \begin{eqnarray*}
   &&\lim_{\Delta t\rightarrow 0}\Big\|\sum\limits_{\texttt{P}:\,Y=\bigcup\limits_iX_i}
          \sum\limits_{\texttt{P}^{'}:\,Y_{X_{i}}=\bigcup\limits_kZ_k}
         (-1)^{|\texttt{P}^{'}|-1}(|\texttt{P}^{'}|-1)!\Big[\frac{1}{\Delta
        t}\times\\
        &&\times\Big[\prod_{Z_k\subset \texttt{P}^{'}}S_{\abs {Z_k}}(-(t+\Delta t),Z_k)
        \prod_{X_i\subset \texttt{P}}g_{\abs {X_i}}(0,X_i)-
        \prod_{Z_k\subset \texttt{P}^{'}}S_{\abs{Z_k}}(-t,Z_k)
        \prod_{X_i\subset \texttt{P}}g_{\abs {X_i}}(0,X_i)\Big]-\\
        &&-\sum\limits_{Z_i\subset P^{'}}(-\mathcal{L}_{|Z_{i}|}(Z_i))
        \prod_{Z_i\subset \texttt{P}^{'}}S_{\abs {Z_i}}(-t,Z_i)
        \prod_{X_i\subset \texttt{P}}g_{\abs
        {X_i}}(0,X_i)\Big]\Big\|=0.
   \end{eqnarray*}
   Indeed, within the context of the corollary of the Liouville theorem and
   the group property of operators $S_{n}(-t)$, $n\geq
1$, (\ref{S(t)}) we obtain
      \begin{eqnarray*}
   &&\lim_{\Delta t\rightarrow 0}\Big\|\sum\limits_{\texttt{P}:\,Y=\bigcup\limits_iX_i}
          \sum\limits_{\texttt{P}^{'}:\,Y_{X_{i}}=\bigcup\limits_kZ_k}
         (-1)^{|\texttt{P}^{'}|-1}(|\texttt{P}^{'}|-1)!\Big[\frac{1}{\Delta
        t}\times\\
        &&\times\Big[\prod_{Z_k\subset \texttt{P}^{'}}S_{\abs {Z_k}}(-(t+\Delta t),Z_k)
        \prod_{X_i\subset \texttt{P}}g_{\abs {X_i}}(0,X_i)-
        \prod_{Z_k\subset \texttt{P}^{'}}S_{\abs{Z_k}}(-t,Z_k)
        \prod_{X_i\subset \texttt{P}}g_{\abs {X_i}}(0,X_i)\Big]-\\
        &&-\sum\limits_{Z_i\subset P^{'}}(-\mathcal{L}_{|Z_{i}|}(Z_i))
        \prod_{Z_i\subset \texttt{P}^{'}}S_{\abs {Z_i}}(-t,Z_i)
        \prod_{X_i\subset \texttt{P}}g_{\abs
        {X_i}}(0,X_i)\Big]\Big\|=\\
                &&=\lim_{\Delta t\rightarrow 0}\Big\|\sum\limits_{\texttt{P}:\,Y=\bigcup\limits_iX_i}
            \sum\limits_{\texttt{P}^{'}:\,Y_{X_{i}}=\bigcup\limits_kZ_k}
        (-1)^{|\texttt{P}^{'}|-1}(|\texttt{P}^{'}|-1)!
       \Big[\frac{1}{\Delta
         t}\Big[
        \prod_{Z_k\subset \texttt{P}^{'}}S_{\abs {Z_k}}(-\Delta t,Z_k)\times\\
         &&\times\prod_{X_i\subset \texttt{P}}g_{\abs {X_i}}(0,X_i)-
        \prod_{X_i\subset \texttt{P}}g_{\abs {X_i}}(0,X_i)\Big]-
        \sum\limits_{Z_i\subset P^{'}}(-\mathcal{L}_{|Z_{i}|}(Z_i))
        \prod_{X_i\subset \texttt{P}}g_{\abs
        {X_i}}(0,X_i)\Big]\Big\|=
         \end{eqnarray*}
        \begin{eqnarray*}
        &&=\int dY\Big|\sum\limits_{\texttt{P}:\,Y=\bigcup\limits_iX_i}
             \sum\limits_{\texttt{P}^{'}:\,Y_{X_{i}}=\bigcup\limits_kZ_k}
        (-1)^{|\texttt{P}^{'}-1}(|\texttt{P}^{'}|-1)!\lim_{\Delta t\rightarrow 0}\Big[\frac{1}{\Delta
         t}\Big[
        \prod_{Z_k\subset \texttt{P}^{'}}S_{\abs {Z_k}}(-\Delta t,Z_k)\times\\
        &&\times\prod_{X_i\subset \texttt{P}}g_{\abs {X_i}}(0,X_i)-
        \prod_{X_i\subset \texttt{P}}g_{\abs {X_i}}(0,X_i)\Big]-
        \sum\limits_{Z_i\subset P^{'}}(-\mathcal{L}_{|Z_{i}|}(Z_i))
        \prod_{X_i\subset \texttt{P}}g_{\abs
        {X_i}}(0,X_i)\Big]\Big|.
   \end{eqnarray*}
\indent The last equality is valid becouse the integrand in this
expression tends to zero as  $\Delta t\rightarrow 0$ uniformly in
$Y$ on any compact set. Therefore, we can pass to the limit as
$\Delta t\rightarrow 0$ in the integral. Hence, equality
(\ref{rozv}) is differentiable in the norm of the
 space $L^{1}_{n}$. Thus, for the initial data from $L^{1}_{n,0}\subset L^{1}_{n}$
  the corresponding Cauchy problem has a unique strong
solution that given by expansions (\ref{1})-(\ref{2}).
\end{proof}
 \section{Properties of the solution  of the nonlinear Liouville hierarchy.}

  We introduce the evolution operator of the solution (\ref{rozv})
\begin{equation}\label{poznach}
     \sum\limits_{\texttt{P}:\,Y=\bigcup\limits_iX_i}
        \mathfrak{A}_{|\texttt{P}|}(t,Y_{X_{i}})
        \prod_{X_i\subset \texttt{P}}g_{\abs {X_i}}(0,X_i)
        \equiv \big(\mathfrak{A}_{t}(g(0))\big)_{|Y|}(Y).
        %\qquad \\
    \end{equation}
 \indent For the evolution operator (\ref{poznach}), the group property is
 valid, i.e.,
       \begin{eqnarray*}
         \big(\mathfrak{A}_{t_{1}}\big(\mathfrak{A}_{t_{2}}(g(0))\big)\big)_{|Y|}(Y)=
         \big(\mathfrak{A}_{t_{2}}\big(\mathfrak{A}_{t_{1}}(g(0))\big)\big)_{|Y|}(Y)=
         \big(\mathfrak{A}_{t_{1}+t_{2}}(g(0))\big)_{|Y|}(Y).
       \end{eqnarray*}
       Indeed, for $g_{n}(0)\in L^{1}_{n}$, $n\geq1$ and for any $t_{1},\,t_{2}\in \mathbb{R}^{1}$,
       according to the notation (\ref{rozv}) and (\ref{kymyl}),
       we have
    \begin{eqnarray*}
      &&\big(\mathfrak{A}_{t_{1}}\big(\mathfrak{A}_{t_{2}}(g(0))\big)\big)_{|Y|}(Y)=\\
      &&\qquad=\sum\limits_{\texttt{P}:\,Y=\bigcup\limits_iX_i}
        \mathfrak{A}_{|\texttt{P}|}(t_{1},Y_{X_{i}})
        \prod_{X_i\subset \texttt{P}}
        \sum\limits_{\texttt{P}^{'}:\,X_{i}=\bigcup\limits_lZ_l}
        \mathfrak{A}_{|\texttt{P}^{'}|}(t_{2},(X_{i})_{Z_{l}})
        \prod_{Z_i\subset \texttt{P}^{'}}g_{\abs
        {Z_l}}(0,Z_l)=\\
        &&\qquad=\sum\limits_{\texttt{P}:\,Y=\bigcup\limits_iX_i}
          \sum\limits_{\texttt{P}_{1}:\,Y_{X_{i}}=\bigcup\limits_kQ_k}
     (-1)^{|\texttt{P}_{1}|-1}(|\texttt{P}_{1}|-1)!
        \prod_{Q_k\subset \texttt{P}^{'}}S_{\abs {Q_k}}(-t_{1},Q_k)
        \prod_{X_i\subset \texttt{P}}\times\\
        &&\qquad\times\sum\limits_{\texttt{P}^{'}:\,X_{i}=\bigcup\limits_lZ_l}
        \sum\limits_{\texttt{P}_{2}:\,(X_{i})_{Z_{l}}=\bigcup\limits_jR_j}
     (-1)^{|\texttt{P}_{2}|-1}(|\texttt{P}_{2}|-1)!
        \prod_{R_k\subset \texttt{P}_{2}}S_{\abs {R_k}}(-t_{2},R_k)
        \prod_{Z_l\subset \texttt{P}^{'}}g_{\abs
        {Z_l}}(0,Z_l).
       \end{eqnarray*}
Having collected the items at identical
           products of the initial data $g_{n}(0)$, $n\geq1$, and taking into account
         the group property of the evolution operators $S_{n}(-t)$, $n\geq1$ (\ref{S(t)}), we obtain
       \begin{eqnarray*}
       &&\big(\mathfrak{A}_{t_{1}}\big(\mathfrak{A}_{t_{2}}(g(0))\big)\big)_{|Y|}(Y)=\\
      &&\qquad=\sum\limits_{\texttt{P}:\,Y=\bigcup\limits_iX_i}
          \sum\limits_{\texttt{P}^{'}:\,Y_{X_{i}}=\bigcup\limits_lZ_l}
     (-1)^{|\texttt{P}^{'}|-1}(|\texttt{P}^{'}|-1)!
        \prod_{Z_l\subset \texttt{P}^{'}}S_{\abs {Z_l}}(-(t_{1}+t_{2}),Z_l)
        \prod_{X_i\subset \texttt{P}}g_{\abs
        {X_i}}(0,X_i)=\\
       &&\qquad=\sum\limits_{\texttt{P}:\,Y=\bigcup\limits_iX_i}
         \mathfrak{A}_{|\texttt{P}|}(t_{1}+t_{2},Y_{X_{i}})
        \prod_{X_i\subset \texttt{P}}g_{\abs{X_i}}(0,X_i)=
        \big(\mathfrak{A}_{t_{1}+t_{2}}(g(0))\big)_{n}(Y).
       \end{eqnarray*}
Similarly,
      \begin{eqnarray*}
         \big(\mathfrak{A}_{t_{2}}\big(\mathfrak{A}_{t_{1}}(g(0))\big)\big)_{n}(Y)=
         \big(\mathfrak{A}_{t_{1}+t_{2}}(g(0))\big)_{n}(Y).
       \end{eqnarray*}
       \indent Let us consider the property of the solution  (\ref{rozv}) for one physically
       motivated example of the initial data that is to say if the initial data for Cauchy problem (\ref{1})-(\ref{2}),
       satisfy the 'chaos' condition \cite{4CGP97}, in other words, the
       sequences of
        correlation functions have the form
       \begin{eqnarray}\label{posl_g(0)}
      g(0)=(1,g_{1}(0,x_{1}),0,0,\ldots),
      \end{eqnarray}
      Indeed, in terms of the sequences $D(0)$, this condition implies
      \cite{4CGP97} that
      \begin{eqnarray*}
     D(0)=\big(1,D_{1}(0,x_{1}), D_{1}(0,x_{1})D_{1}(0,x_{2}),\ldots\big),
     \end{eqnarray*}
 which implies that particle distributions are statistically
 independent at the initial time instant.
  Making use of the relation (\ref{g(D)}), we obtain the initial
  condition (\ref{posl_g(0)}) for the correlation functions.\\
\indent For the initial data (\ref{posl_g(0)}), the formula for the
solution (\ref{rozv}) of the initial-value problem
(\ref{1})-(\ref{2}) is simplified and reduces to
       \begin{equation}\label{rozv_Chaos}
        g_{n}(t,x_{1},\ldots,x_{n})=\mathfrak{A}_{n}(t,x_{1},\ldots,x_{n})
        \prod_{i=1}^{n}g_{1}(0,x_i).
              \end{equation}
             In this case the following estimate holds
          \begin{eqnarray*}
          \|g_{n}(t)\|\leq n!e^{n+1}\|g_{1}(0)\|^{n}.
          \end{eqnarray*}
\indent In the case of the initial data (\ref{posl_g(0)}), the
solution (\ref{rozv_Chaos}) of the Cauchy problem for the nonlinear
Liouville hierarchy may be rewritten in a different form. If $n=1$,
we have
        \begin{equation*}
        g_{1}(t,x_{1})=\mathfrak{A}_{1}(t,x_{1})g_{1}(0,x_1)=g_{1}(0,q_1-p_{1}t,p_{1})
                \end{equation*}
       i.e., an explicit expression that describes the evolution of one particle.
       Then, within the context of the definition of the $1$-st order cumulant,
       $\mathfrak{A}_{1}(t)$, and inverse to it evolution operator
$\mathfrak{A}_{1}(-t)$, we can express the correlation functions
$g_{n}(t)$, $n\geq2$, in terms of the one-particle correlation
function $g_{1}(t)$ making use of formula
(\ref{rozv_Chaos}).\\
 Finally, formula (\ref{rozv_Chaos}) for $n\geq2$ is given by
         \begin{equation*}
        g_{n}(t,x_{1},\ldots,x_{n})=\widehat{\mathfrak{A}}_{n}(t,x_{1},\ldots,x_{n})
        \prod_{i=1}^{n}g_{1}(t,x_i),
              \end{equation*}
              where $\widehat{\mathfrak{A}}_{n}(t,x_{1},\ldots,x_{n})$
              is the $n$-th order
              cumulant of the scattering operators
              $\widehat{S}_{t}$, i.e.,
              \begin{equation*}
      \widehat{S}_{t}(x_{1},\ldots,x_{n})=
      S_{n}(-t,x_{1},\ldots,x_{n})\prod_{i=1}^{n}S_{1}(t,x_{i}),
           \end{equation*}
          whose generators are determined by the operator
           $\mathcal{L}^{int}(\ref{liyvilnian})$ in terms of the interaction potential of the system.
\section*{Acknowledgement}

 The author is pleased to thank  Prof.~Victor Gerasimenko for the formulation of the problem
 and for
many useful discussions.


\begin{thebibliography}{99}
\bibitem{1BJinr}
        {\sl Bogolyubov N.} On the stochastic processes in dynamical
        systems.--- in EPAN, Dubna: JINR, {\bf 9}, N~4, 1978.
\bibitem{2Green}
        {\sl  Green M.S.} Boltzmann equation from the
               statistical mechanical point of view.---
               Journal Chem. Phys., {\bf 25}, N~5, 1956, pp.~836-855.
\bibitem{3GR02}
        {\sl Gerasimenko V.I. and Ryabukha T.V.,}
        Cumulant representation of solutions of the BBGKY
        hierarchy of equations,
        {\it Ukr. Math. J.}.~-- 2002~-- {\bf 54}, N~10.~--- pp.~1583--1601.
\bibitem{4CGP97}
        {\sl Cercignani C., Gerasimenko V.I., Petrina D.Ya.}
        Many-Particle Dynamics and Kinetic Equations.~---
        Kluwer Acad. Publ., 1997.~--- p.~256~.
\bibitem{5GRS04}
        {\sl Gerasimenko V.~I., Ryabukha T.~V., Stashenko M.~O.}
        On the structure of expansions for the BBGKY Hierarchy Solutions~//
        J. Phys. A: Math. Gen.~-- 2004.~-- {\bf 37}.~--- pp.~9861-9872.
\bibitem{6PGM02}
        {\sl Petrina D.Ya., Gerasimenko V.I., Malyshev P.V.}
        Mathematical Foundations of Classical Statistical Mechanics.
        Continuous Systems.~---  London and N.Y.: Taylor \& Francis Inc. (Second ed.), 2002.~---
        p.~352.
\bibitem{7B_Lec_49}{ Bogolyubov N.} Lectures on Quantum
               Statistics.--- Kyiv: Rad. shkola, 1949.
\end{thebibliography}
\end{document}